\documentclass[useAMS,usenatbib,referee]{biom}
\usepackage{lscape}
\usepackage[utf8]{inputenc}
\def\bSig\mathbf{\Sigma}

\usepackage{color,graphics}
\usepackage{graphicx}
\usepackage[figuresright]{rotating}
\graphicspath{{./Figures/}}
\usepackage{amsmath}
\newcommand\numberthis{\addtocounter{equation}{1}\tag{\theequation}}
\usepackage{amsfonts}
\DeclareMathOperator{\e}{e}
\usepackage{amssymb}

\title[Joint latent class model for longitudinal data and interval-censored semi-competing events]{Joint latent class model for longitudinal data and interval-censored semi-competing events: Application to dementia}

	   \author{Anaïs Rouanet$^{1,2,*}$\email{anais.rouanet@isped.u-bordeaux2.fr}, Pierre Joly$^{1,2}$, Jean-François Dartigues$^{1,2}$, \\
	   \textbf{Cécile Proust-Lima$^{\bm{1,2}}$  and
Hélène Jacqmin-Gadda$^{\bm{1,2}}$ } \\
$^{1}$INSERM, Centre INSERM U897 - Epidemiologie - Biostatistiques, F-33076 Bordeaux, France\\
$^{2}$Université de Bordeaux, ISPED, 146 rue Léo Saignat, F-33076 Bordeaux, France }
\begin{document}

\date{{\it Received October} 2004. {\it Revised February} 2015.\newline
{\it Accepted March} 2005.}

\pagerange{\pageref{firstpage}--\pageref{lastpage}} \pubyear{2015}

\volume{59}
\artmonth{February}
\doi{10.1111/j.1541-0420.2005.00454.x}


\label{firstpage}


\begin{abstract}
Joint models are used in ageing studies to investigate the association between longitudinal markers and a time-to-event, and have been extended to multiple markers and/or competing risks. The competing risk of death must be considered in the elderly because death and dementia have common risk factors. Moreover, in cohort studies, time-to-dementia is interval-censored because dementia is only assessed intermittently. So subjects can become demented and die between two follow-up visits without being diagnosed. To study pre-dementia cognitive decline, we propose a joint latent class model combining a (possibly multivariate) mixed model and an illness-death model handling both interval censoring (by accounting for a possible unobserved transition to dementia) and semi-competing risks. Parameters are estimated by maximum likelihood handling interval censoring. The correlation between the marker and the times-to-events is captured by latent classes, homogeneous groups with specific risks of death and dementia and profiles of cognitive decline. We propose markovian and semi-markovian versions. Both approaches are compared to a joint latent class model for standard competing risks through a simulation study, and then applied in a prospective cohort study of cerebral and functional ageing to distinguish different profiles of cognitive decline associated with risks of dementia and death. The comparison highlights that among demented subjects, mortality depends more on age than duration of dementia. This model distinguishes the so-called terminal pre-death decline (among non-demented subjects) from the pre-dementia decline.

\end{abstract}
%
%


\maketitle

\section{Introduction}
\label{s:intro}

Joint models are becoming increasingly popular as they allow an analysis of the association between the risk of an event and the change over time of a longitudinal marker \citep{tsiatis,rizopoulos}. In a cognitive ageing study, the link between cognitive decline and dementia needs to be understood to better describe the course of the disease in the pre-diagnostic stage, and to develop  prediction tools for the risk of dementia. Moreover, modeling the evolution of cognitive markers without modeling jointly the risk of dementia may lead to biased estimations  of the change over time of the marker as collection of cognitive measures is often stopped after dementia onset, inducing non-random missing data. 
Joint models correct for this bias  by accounting for the association between the marker and the time-to-event.\\

\cite{Wulfsohn} proposed shared random effects models where a function of the random effects from the longitudinal model is included in the survival model, thus capturing the correlation between the time-to-event and the marker. The risk of the event is then partly explained by the individual dynamics of the marker trajectory. An alternative is the joint latent class mixed model, developed by \cite{Lin} which considers a heterogeneous population, divisible into several homogeneous latent subgroups, with a specific risk of the event and a specific evolution of the marker. A significant computational advantage is the replacement of the integrals on the random effects by a sum over the classes, such that the likelihood of this model has a closed form.\\

When studying the risk factors or natural history of Alzheimer's disease, the most frequent cause of dementia in the elderly, it is important to account for the competing risk of death as dementia and death have common risk factors. \cite{Elashoff} and \cite{Williamson} proposed a shared random effects model accounting for competing risks and \cite{Proust_2015} developed a joint latent class model for competing risks to study multiple longitudinal markers of different natures.
However, studies of the risk of dementia are made more difficult by the interval censoring of time-to-dementia. Indeed, in cohort studies, patients are observed intermittently and the age at dementia onset is not precisely known as dementia can only be diagnosed at clinical follow-up visits. As such, there is an interval of uncertainty between the last visit where the patient has been seen to be healthy and the visit where a diagnosis was made. More importantly, a patient can become demented and die between two visits without being diagnosed as demented. Consequently, the risk of dementia may be underestimated when interval censoring is not accounted for, for example when considering only the first observed event in the standard competing risks model. \cite{joly_2002} proposed an illness-death model to fix this issue but this has not yet been implemented in a joint model. Death and dementia are semi-competing events since dementia can not occur after death but death occurs after dementia.\\

To our knowledge, only one joint model combining a multi-state model and a mixed model for a longitudinal marker has previously been proposed. Within the framework of the shared random effects approach, Dantan et al. (2011) described a joint model combining a two-phase mixed model with a random change-point and a multi-state model. The underlying clinical idea was an acceleration of the cognitive decline before dementia onset, which was modelled by a second phase with a different slope in the mixed model. In this model, interval censoring was not a critical issue as death depended on the current value of the marker and not on the current state.\\ 

In this work, we propose a joint latent class illness-death model for semi-competing interval-censored events and a longitudinal marker. We propose two versions of the model, a markovian and a semi-markovian version, and an extension for the joint analysis of multiple longitudinal markers. In the following section, we detail the model and the estimation procedure. In section 3, we present a simulation study to evaluate the estimation procedure and compare the proposed approach with a joint latent class model for competing risks that does not take into account interval censoring. In section 4, the model is applied to study cognitive decline before Alzheimer's disease diagnosis and death using data from the French Paquid cohort, including 3,777 subjects followed over 20 years with regular cognitive evaluation.
\section[]{Methods}
\subsection{Notations}
Let $Y_{ij}$ denote the score of the psychometric test for subject $i$, for $i=1,...,N$, at time $t_{ij}$, for $j=1,...,n_i$. We denote by $T^A_i$ the age at dementia onset and $T^D_i$ the age at death. We assume that age at dementia onset is interval-censored while age at death is only right-censored as generally, exact ages of death are collected in cohort studies. Thus, $D_i $ denotes the vector of collected variables about the event: $ D_i=({T_0}_i,L_i, R_{i},\delta^A_{i},T_i,\delta^D_{i})^{\top}$ where ${T_0}_i$ is the age at inclusion, $L_{i}$ is the age at the last visit where the subject has been seen to be healthy, $R_{i}$ is the age at the visit of diagnosis if the subject is diagnosed with dementia ($R_i=+\infty$ if not diagnosed), $T_i$ is the age at death or at the end of the follow-up,  $\delta^A_{i}$ is the indicator of Alzheimer's diagnosis or dementia ($\delta^A_i=1$ if $R_i \leq T_i$ and 0 otherwise) and $\delta^D_{i}$ is the indicator of death ($\delta^D_i=1$ if $T^D_i=T_i$ and 0 otherwise).

\subsection{Joint latent class illness-death model}

\begin{figure}[h!]
\begin{center}
\includegraphics[scale=0.5]{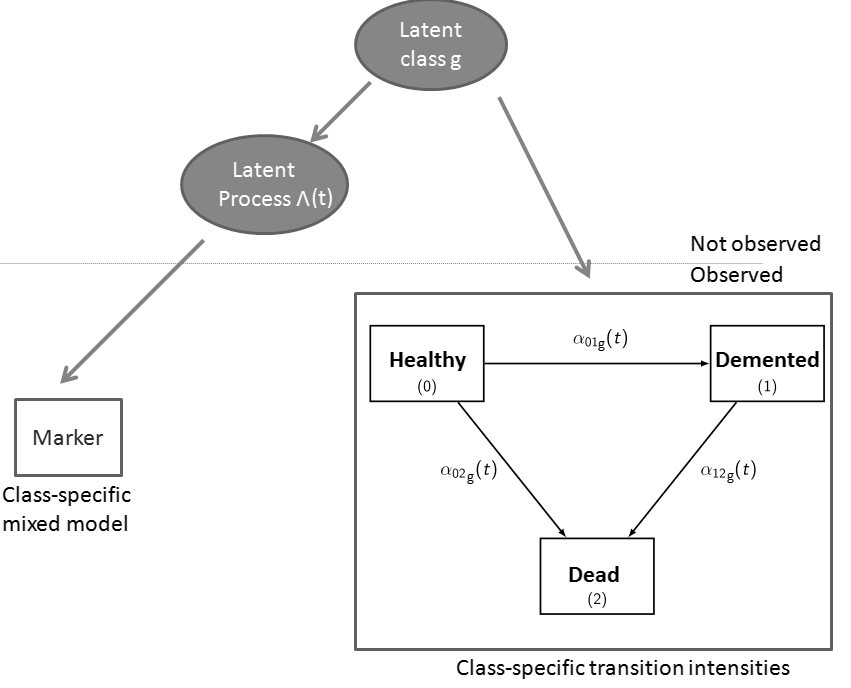}\\
\caption{Joint latent class illness-death model: the latent classes correspond to homogeneous subgroups of subjects with a specific marker trajectory and specific transition intensities to dementia and death.}
\label{fig::model}
\end{center}
\end{figure}

The model relies on the hypothesis that the population is heterogeneous and can be divided into G homogeneous latent classes. Each class has specific transition intensities for dementia and death and a specific marker trajectory, as displayed in  Figure \ref{fig::model}. A central assumption states that the marker and the times-to-events are independent conditionally on the classes and covariates.  \\

We first describe the probability $\pi_{ig}$, for subject $i$, of belonging to class $g$, for $g=1,...,G$, with a multinomial logistic model:
\begin{equation}
\pi_{ig}= P(c_i=g)= \frac{\rm exp({X^{\top}_{pi}}~ \zeta_g)}{\sum^G_{m=1} \rm exp({X^{\top}_{pi}}~ \zeta_m)}
\end{equation}
The latent class membership variable $c_i$ is  $c_i=g$ if subject $i$ belongs to class $g$ and $X_{pi}$ is a vector of covariates. We choose class G as the reference class so that $\zeta_G = 0$ to ensure identifiability.\\

We denote by $\Lambda(\cdot)$ the latent process which stands for the true cognitive level. The conditional distribution of $\Lambda(t)$ given the latent class is defined by a standard linear mixed model, without residual error, with class-specific parameters:
\begin{equation}
\Lambda_i(t_{ij})= {X^{\top}_{ij}}~ \beta_{g}+{Z^{\top}_{ij}}~ u_{ig},
\end{equation}
where $X_{ij}$ is a vector of covariates of subject $i$ at time $t_{ij}$, $\beta_{g}$ is the vector of class-specific regression parameters and $Z_{ij}$ is a sub-vector of $X_{ij}$. The random effects $u_{ig} \sim \mathcal{N}(0,\sigma^2_g B)$ have proportional variance matrices over the classes, with $\sigma^2_G=1$. We denote by $U$ the Cholesky transformation of the matrix $B$, which is a lower triangular matrix satisfying $UU^{\top}=B$. The marker $Y_{ij}$ is considered as a measure with error of the latent process $\Lambda_i(t_{ij})$ at time $t_{ij}$:
\begin{equation}
Y_{ij}= \Lambda_{i}(t_{ij}) +\epsilon_{ij}  \mbox{  with  }\epsilon_{ij} \sim \mathcal{N}(0,\sigma^2_e).
\end{equation}

The transition intensities of dementia and death are modelled simultaneously using an illness-death model with three class-specific transition intensities (Figure \ref{fig::model}). In this way, we can distinguish the transition intensities of death for healthy and demented subjects. We propose both a markovian and a semi-markovian model.

\subsection{Markovian model}

Given the latent class g, the transition intensity from state $k$ to state $l$ depends on age $t$ and it is modelled by a proportional hazards model with class-specific parameters:
\begin{equation}
\alpha_{klig}(t) = \alpha^0_{klg}(t)~ \rm{\exp}(W^{\top}_{kli}~ \gamma_{klg}),
\end{equation}
where $\alpha^0_{klg}$ is the baseline transition intensity, $W^{\top}_{kli}$ is a vector of covariates  and $\gamma_{klg}$ are class-specific regression parameters. In the following paragraphs, we will refer to the cumulative transition intensities using the notation: $\displaystyle{A_{klg}(t) = \int_0^t \alpha_{klg}(s) ds}$.

\subsection{Semi-markovian model}
Alternatively, the transition intensity of death among demented subjects may depend on the time spent in the dementia state instead on age, leading to a semi-markovian illness-death model:
\begin{equation}
\alpha_{12ig}(t,T^A_i)= \alpha_{12ig}(t-T^A_i)= \alpha^0_{12g}(t-T^A_i)~ \rm{\exp}(W^{\top}_{12i}~\gamma_{12g}),
\end{equation}
where $T^A_i$ is the age at dementia onset so $t-T^A_i$ is the time spent in the dementia state.

\subsection{Log-likelihood of the markovian model}
\begin{figure}[h!]
\begin{center}
\includegraphics[scale=0.5]{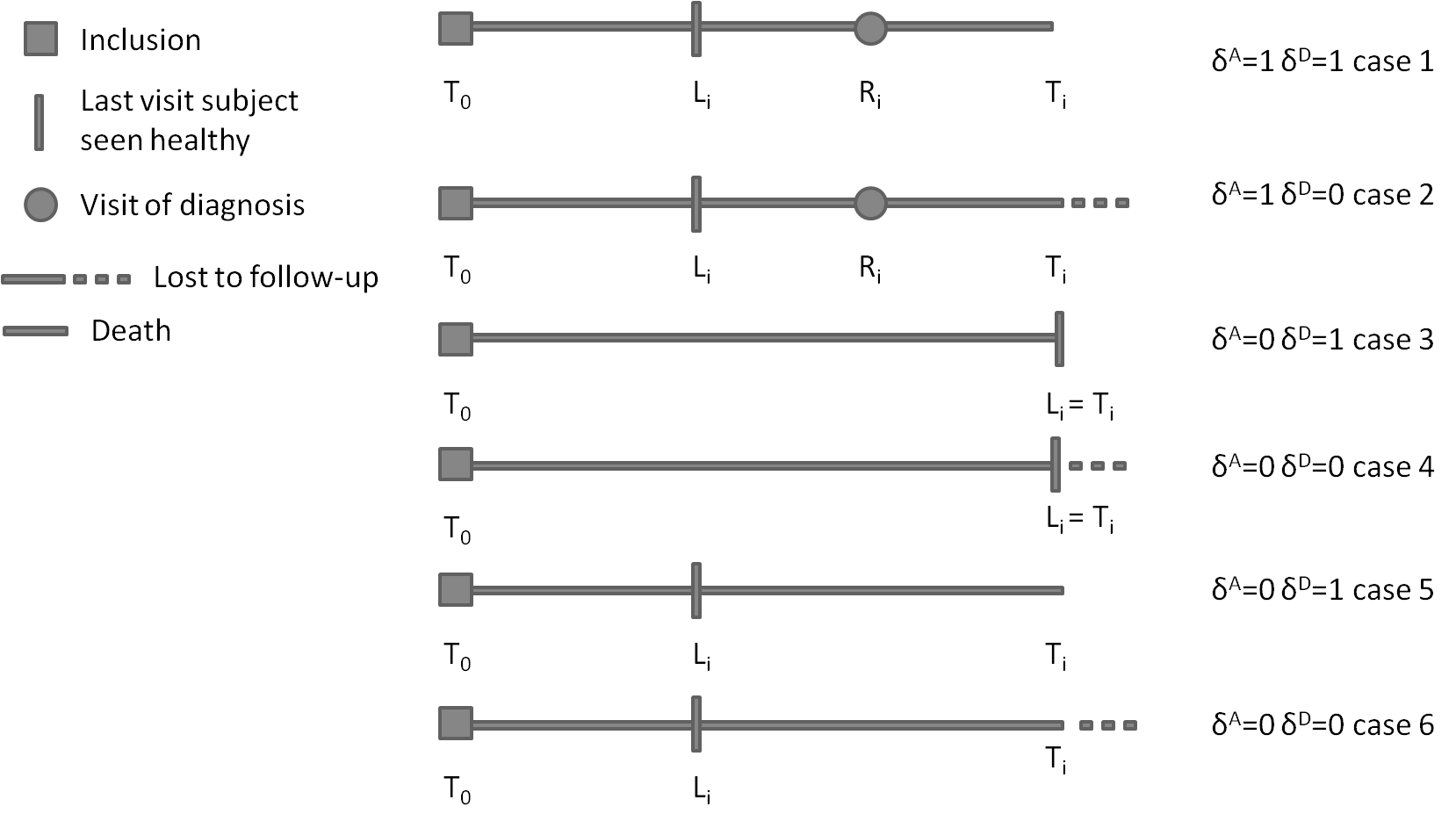}
\caption{Possible observation patterns for dementia and death. To obtain a more flexible program, we also implemented the computation of the likelihood for subjects with an exact date of dementia onset ($L_i=R_i$) although this date was never known in our dataset.}
\label{cens}
\end{center}
\end{figure}

Let $\theta_G$ denote the vector including the regression, variance and  baseline transition intensities parameters. The contribution $\mathcal{L}_i$ of any subject $i$ to the global log-likelihood $\mathcal{L}(\theta_G)$ is the weighted sum of his or her contributions over the G classes. According to the conditional independence assumption, the individual conditional contribution to the likelihood given the class is the product of the conditional contributions of the mixed model and of the multi-state model, as follows:
\begin{align*}
\mathcal{L}(\theta_G)=\sum\limits_{i=1}^{N}\mathcal{L}_i
= \sum\limits_{i=1}^{N} \log \big[  \sum\limits_{g=1}^{G} \pi_{ig} &f(Y_i|c_i=g;\theta_G)P(D_i|c_i=g;\theta_G) \big]\\
& - \sum\limits_{i=1}^{N} \log \big[  \sum\limits_{g=1}^{G} \pi_{ig}\e^{-A_{01ig}(T_{0i};\theta_G)-A_{02ig}(T_{0i};\theta_G)} \big]\numberthis
   \label{eq::lk}
\end{align*}
where $f(Y_i|c_i=g;\theta_G)$ is a multivariate gaussian density with mean $E_{ig} = X^{\top}_{i} \beta_g$ and variance matrix $V_{ig} = \sigma^2_g~Z_{i} BZ^{\top}_{i} + \sigma^2_e I$ and $X_i$ and $Z_i$ are design matrices with row vectors $X^{\top}_{ij}$ and $Z^{\top}_{ij}$. Then, $P(D_i|c_i=g;\theta_G)$ is detailed below for each possible observation pattern for death and dementia (as illustrated in Figure \ref{cens}). The second part accounts for the delayed entry, representing the probability of being alive and healthy at entry, which is the condition for inclusion in the sample.

\begin{itemize}
\item Subject diagnosed as demented (cases 1 and 2, Figure \ref{cens}):
\end{itemize}
$$P^d_{ig}({T_0}_i,L_{i}, R_{i},1,T_i,\delta^D_{i};\theta_G)= \displaystyle{\int_{L_{i}}^{R_{i}}} \e^{-A_{01ig}(u)}\e^{-A_{02ig}(u)} \alpha_{01ig}(u) \e^{-(A_{12ig}(T_{i})-A_{12ig}(u))} {\alpha_{12ig}(T_i)}^{\delta^D_i} \mathrm du  $$
The subject remained healthy and alive until age $u$ between $L_i$ and $R_i$, became demented at $u$, remained alive until $T_i$ and possibly died at $T_i$ (if $\delta^D_i=1$).

\begin{itemize}
\item Subject observed healthy at the end of the follow-up (cases 3 and 4, Figure \ref{cens}):
\end{itemize}
$$P^{h}_{ig}({T_0}_i,L_{i}, R_{i},0,T_i,\delta^D_{i};\theta_G)=\e^{-A_{01ig}(T_i)-A_{02ig}(T_i)}{\alpha_{02ig}(T_i)}^{\delta^D_i}$$
The subject remained healthy and alive until $T_i$ and possibly died at $T_i$. Case 4 corresponds to subjects healthy at the last visit and with no information on vital status after this visit. Case 3 is not observed in the Paquid study because subjects never die the very day of the visit. Consequently, we can never be totally sure that a subject who died was free of dementia. Nevertheless, this case may be observed for other pathologies.

\begin{itemize}
\item Subject with unknown dementia status at the end of the follow-up (cases 5 and 6, Figure \ref{cens}):
\end{itemize}
\begin{align*}
P^u_{ig}({T_0}_i,L_{i}, R_{i},0,&T_i,\delta^D_{i};\theta_G)=\e^{-A_{01ig}(T_i)-A_{02ig}(T_i)}{\alpha_{02ig}(T_i)}^{\delta^D_i}\\
&+ \displaystyle{\int_{L_{i}}^{T_i}} \e^{-A_{01ig}(u)-A_{02ig}(u)} \alpha_{01ig}(u) \e^{-(A_{12ig}(T_i)-A_{12ig}(u))} {\alpha_{12ig} (T_i)}^{\delta^D_i}\mathrm du
\end{align*}
The numerator of $P^u_{ig}$ accounts for the two possible trajectories: either the subject remained healthy until the end of the follow-up $T_i$ (and possibly died), or he/she became demented between the last visit $L_i$, where he/she was observed healthy, and $T_i$. If so, the subject remained healthy and alive until age $u$ between $L_i$ and $T_i$, became demented at $u$, remained alive until $T_i$ and was lost to follow-up or died at $T_i$. 


\subsection{Likelihood of the semi-markovian model}
In the case of a semi-markovian model, the individual conditional contribution to the likelihood of the multi-state model is changed for the observation patterns 1, 2, 5 and 6 as follows:

\begin{itemize}
\item Subject diagnosed as demented (cases 1 and 2, Figure \ref{cens}):
\end{itemize}
$$P^d_{ig}({T_0}_i,L_i, R_i,1,T_i,\delta^D_{i};\theta_G)=\displaystyle{\int_{L_i}^{R_i}} \e^{-A_{01ig}(u)-A_{02ig}(u)} \alpha_{01ig}(u) \e^{-A_{12ig}(T_i-u)}{\alpha_{12ig}(T_i-u)}^{\delta^D_i} \mathrm du $$

\begin{itemize}
\item Subject with unknown dementia status at the end of follow-up $T_i$ (cases 5 and 6, Figure \ref{cens}):
\end{itemize}
\begin{align*}
P^u_{ig}({T_0}_i,L_{i}, R_{i},0,T_i,\delta^D_{i};\theta_G)=&\e^{-A_{01ig}(T_i)-A_{02ig}(T_i)}{\alpha_{02ig}(T_i)}^{\delta^D_i}\\
 &+ \displaystyle{\int_{L_{i}}^{T_i}} \e^{-A_{01ig}(u)-A_{02ig}(u)} \alpha_{01ig}(u)
 \e^{-A_{12ig}(T_i-u)}{\alpha_{12ig}(T_i-u)}^{\delta^D_i}\mathrm du 
\end{align*}

This model assumes that the transition intensity of death among demented subjects depends on the duration of dementia and not on age.

\subsection{Optimisation algorithm}
The maximum likelihood estimators are obtained for a fixed number of classes G by a Newton-Raphson-like algorithm \citep{marq98}. If necessary, at each iteration $p$, the Hessian matrix $H^{(p)}$ is diagonal-inflated to obtain a positive definite matrix $H^{*(p)}$. The vector of parameters is then updated by $\theta_G^{(p+1)}=\theta_G^{(p)} - \kappa~ U(\theta_G^{(p)})^{\top}[H^{*(p)}]^{-1}U(\theta_G^{(p)})$ with $U(\theta_G^{(p)})$ the gradient at iteration $p$ and $\kappa$ the improvement control parameter, optimized using a line search strategy. The convergence criteria are reached when the change in the likelihood, the change in the estimates between two iterations and the criterion $U(\theta_G^{(p)})^{\top}[H^{(p)}]^{-1}U(\theta_G^{(p)})$ are less than $10^{-3}$, $10^{-3}$ and $10^{-2}$ respectively. The variances of the estimates are obtained with the inverse of $H^{(p)}$. For each value of $G$, the estimation process is repeated with different initial values to insure convergence. Finally, the number of classes G is chosen by minimising the Bayesian Information Criterion (BIC) \citep{BIC}, which favors parsimonious models and is recommended for mixture models \citep{Hawkins}.

\subsection{Extension to multiple non-gaussian markers}
The model can be extended to the analysis of K non-gaussian markers as in \cite{CSDA}. In psychometrics, quantitative tests are frequent, with asymmetric distributions and ceiling or floor effects. Moreover, as these tests are highly correlated, they  may be considered as measures with error of  a common latent process that stands for the true latent cognitive level underlying the various tests.
A parametric monotonic function $\Psi_k(\cdot,\eta_k)$, such as a Beta cumulative distribution function or a spline function, with $\eta_k$ a vector of test-specific parameters, can then be used to model the link between the observed markers $Y_{ijk}$ and the latent process $\Lambda(t_{ijk})$:
\begin{equation} \label{eq::psi}
\Psi_k(Y_{ijk};\eta_k)= \Lambda_{i}(t_{ijk})+X^{k\top}_{ij}~  \beta^k+\epsilon_{ijk} \mbox{ with } \epsilon_{ijk} \sim \mathcal{N}(0,\sigma^2_{e_k})
\end{equation}
Note that for more flexibility, the model (\ref{eq::psi}) may also include marker-specific effects $\beta^k$ of some covariates $X^k_{ij}$. Consequently, we can define transformed scores $\tilde{Y}_{ijk}$ for test $k$ and subject $i$ at time $t_{ijk}$, for $i=1,...,N, j=1,...,n_{ik}, k=1,...,K$, on the scale of the latent process:
\begin{equation}
\tilde{Y}_{ijk}= \Psi_k(Y_{ijk};\eta_k)
\end{equation}

The log-likelihood is then defined by:
 \begin{align*}
\mathcal{L}(\theta_G)
&= \sum\limits_{i=1}^{N} \log \big( \sum\limits_{g=1}^{G} \pi_{ig}  \Phi_{g}(\widetilde{Y_i}|c_i=g;\theta_G)~\Big[ \prod^K_{k=1} \prod^{n_{ik}}_{j=1} J(\Psi_k(Y_{ijk};\eta_k)) \Big] ~ P(D_i|c_i=g;\theta_G) \big)\\
&= \sum\limits_{i=1}^{N} \log \Big[ \sum\limits_{g=1}^{G} \pi_{ig}  \Phi_{g}(\widetilde{Y_i}|c_i=g;\theta_G)~ P(D_i|c_i=g;\theta_G) \Big]+\sum\limits_{i=1}^{N} \sum^K_{k=1} \sum^{n_{ik}}_{j=1} \log \Big[J(\Psi_k(Y_{ijk};\eta_k)) \Big],
\end{align*}
where $J$ is the Jacobian, 
$\Phi_{g}(\widetilde{Y_i}|c_i=g;\theta_G)$ is a multivariate gaussian density with mean $E_{ig} = (E^{\top}_{i1g},...,E^{\top}_{iKg})^{\top}$ and the elements of $E_{ikg}$ are $E_{ijkg} = X^{\top}_{i}(t_{ijk})  ~\beta_g + X^{k\top}_{ij}  ~\beta^k$, with $X^{\top}_{i}(t_{ijk})$ is the vector of covariate values included in $X_i$ at time $t_{ijk}$, and variance matrix:

\begin{align*}
V_i=&\begin{pmatrix} Z_i^1 \\ \vdots \\ Z_i^K \end{pmatrix}
        \sigma^2_g B ~~ \left( Z_i^{1\top} ~~\mbox{...} ~~ Z_i^{K\top} \right) + \begin{pmatrix} \sigma^2_{e_1} I_{n_{i1}}& 0 & 0\\
        0 &\ddots & 0\\
        0 & 0 &  \sigma^2_{e_K} I_{n_{iK}} \end{pmatrix}
\end{align*}


\subsection{Goodness-of-fit }

Once the parameters are estimated, we can compute the posterior probability of belonging to class $g$,
\begin{equation}
 P(c_i=g|Y_i,D_i;\hat{\theta}_G) = \frac{\hat{\pi}_{ig}~ f(Y_i|c_i=g;\hat{\theta}_G)~P(D_i|c_i=g;\hat{\theta}_G)}{\vphantom{\int^T} \sum^G_{m=1} \hat{\pi}_{im}f(Y_i|c_i=m;\hat{\theta}_G)~P(D_i|c_i=m;\hat{\theta}_G)},\\
 \end{equation}
  and subjects are assigned to the class with the highest probability. First, we propose to assess the goodness-of-fit of the longitudinal predictions, conditional on the classes. To do so, we split the timescale into five-year age groups $[\tau_{q},\tau_{q+1}]$. For each class $g$, we then compare the class-specific predicted mean evolution of the marker, weighted by the posterior class membership probability
\begin{equation*}
\hat{\mu}_{gq}=\frac{\sum_{(i,j)|\tau_q<t_{ij}<\tau_{q+1}} E(Y_{ij}|c_i=g;\hat{\theta}_G)~P(c_i=g|Y_i,D_i;\hat{\theta}_G)}{\vphantom{\int^T}\sum_{(i,j)|\tau_q<t_{ij}<\tau_{q+1}}~ P(c_i=g|Y_i,D_i;\hat{\theta}_G) },
\end{equation*}
 to the observed mean evolution weighted by the same posterior probability:
\begin{equation*}
\hat{\mu}^o_{gq}=\frac{\sum_{(i,j)|\tau_q<t_{ij}<\tau_{q+1}} Y_{ij}~P(c_i=g|Y_i,D_i;\hat{\theta}_G)}{\vphantom{\int^T} \sum_{(i,j)|\tau_q<t_{ij}<\tau_{q+1}}~ P(c_i=g|Y_i,D_i;\hat{\theta}_G) }
\end{equation*}

The assessment can also be done conditionally on the random effects, comparing $\hat{\mu}^o_{gq}$ to
 \begin{equation*}
\hat{\mu}_{gq}^{u}=\frac{\sum_{(i,j)|\tau_q<t_{ij}<\tau_{q+1}} E(Y_{ij}|c_i=g,\hat{u}_{ig};\hat{\theta}_G)~P(c_i=g|Y_i,D_i;\hat{\theta}_G)}{\vphantom{\int^T} \sum_{(i,j)|\tau_q<t_{ij}<\tau_{q+1}}~ P(c_i=g|Y_i,D_i;\hat{\theta}_G) }
 \end{equation*}
with $ \hat{u}_{ig} =  E(u_i|Y_i,c_i=g;\hat{\theta}_G)$ the bayesian estimates of the random effects given the class g.\\


Secondly, to evaluate the goodness-of-fit of the parametric illness-death predictions conditionally on the classes, we compare the predicted class-specific cumulative incidences of the three transitions to the class-specific predictions obtained by a semi-parametric illness-death model \citep{touraine}. Each transition intensity of this model is modelled by a proportional hazards model with baseline transition intensities modelled by M-splines and estimated by penalized likelihood. The contribution to the likelihood of any subject $i$ is weighted by the individual posterior probability $P(c_i=g|Y_i,D_i;\hat{\theta}_G)$ obtained by the joint latent class illness-death model. Note that the cumulative incidences for transitions 0-1 and 0-2 are estimated given that the subject is alive and healthy at age 65, and the cumulative incidence for transition 1-2 is estimated given that the demented subject is alive at age 65, as follows:

\begin{align*}
F_{0lg}(t)&=\frac{\displaystyle{\int_{65}^t} \e^{-A_{01g}(u)-A_{02g}(u)}\alpha_{0lg}(u) du}{\vphantom{\int^T}\e^{-A_{01g}(65)-A_{02g}(65)}}, ~~ l=1,2\\
F_{12g}(t)&=\frac{\displaystyle{\int_{65}^t} \e^{-A_{12g}(u)}\alpha_{12g}(u) du}{\vphantom{\int^T}\e^{-A_{12g}(65)}}
  \end{align*}

\section{Simulations}
\subsection{Design}
We carried out simulations in order to evaluate the estimation procedure and compare the estimations with those obtained using a joint latent class model for competing events without accounting for interval censoring.\\

Data were generated with a model with 2 latent classes with probability  $\pi_1=0.5$ ($\zeta_1=0$) for the first class. Of note, $\pi_{1}$ represents the probability of belonging to the first class in the general population. As we only include in the cohort subjects who were healthy and alive at the first visit, we introduce a selection bias. The proportion of each class in the selected sample may then be different.\\

For each subject, age at entry is generated from a uniform distribution on [65,85]. Age at dementia onset and age at death (for demented and non-demented subjects) are generated from Weibull distributions with class-specific parameters (shape parameter $\lambda^{(1)}_{klg}$ and scale parameter $\lambda^{(2)}_{klg}$ for transition intensity from state $k$ to state $l$ in class $g$). The transition intensities account for a common effect of a binary covariate X, also generated from a binomial distribution with parameter 0.5:
\begin{equation}
\alpha_{klig}(t)  = \lambda^{(1)}_{klg}~ {\lambda^{(2)}_{klg}}^{\lambda^{(1)}_{klg}} ~ t^{(\lambda^{(1)}_{klg}-1)}~ \e^{\gamma_{kl} X_i} \mbox{ for } k=0,1;~ l=1,2;~ g=1,2
\end{equation}

The scores of the psychometric test are generated by a linear mixed model including fixed and random effects on the intercept and the slope, with an adjustment for the covariate X, common over the classes:
  \begin{equation}
  Y_{ij}  = \beta_{0g} + \beta_{1g} ~ t_{ij} + u_{ig}^{(0)} + u_{ig}^{(1)} ~ t_{ij} + \beta_X ~ X_i,    ~~~~ u_{ig} \sim \mathcal{N}(0,\sigma^2_g B)
  \end{equation}
Time is a linear transformation of age: $t=\frac{age-65}{10} $. No parametric transformation $\Psi$ is used so the generated scores are considered as the observed ones. Two designs of follow-up were generated: the follow-up visits were scheduled either every 2 or 4 years from inclusion to the minimum between the visit following dementia onset, death or the administrative right-censoring which is 20 years after inclusion. For each design, we generated 500 samples of 500 subjects. The simulated parameters were similar to the ones obtained by a joint markovian illness-death model with two latent classes on the Paquid dataset, without any linear transformation.\\

We define the age at diagnosis as the age at the first visit following  dementia onset if the generated age at death is old enough. Subjects who die before the next visit are considered as censored for dementia at the last visit before dementia onset.\\

 On average, over the 500 samples simulated within the "visits every two years" framework, 23\% of the subjects were observed demented,  17.7\% died after dementia diagnosis and 62\% died without dementia diagnosis. An average of 33.8\% were allocated to the first class, of which 33.8\% were observed demented, 65.3\% died with no dementia diagnosis and 30.9\% died after  dementia diagnosis. In the second class, 17.9\% were seen demented,  60.5\% died with no dementia diagnosis and 10.9\% died after dementia diagnosis.\\

We also estimated a joint latent class model for competing risks, which does not account for  interval censoring, on the same simulated data. In the standard competing risks framework, the outcome is a couple ($T$, $\delta$) with $T$ the time to the first event or the censoring time and $\delta$ the indicator of the cause of the first event ($\delta=1$ if demented, $\delta=2$ if dead and $\delta=0$ if censored). The estimation of this model on interval-censored data requires to impute this couple ($T$, $\delta$) for some subjects. If the subject is observed demented before death, the recorded transition is the health-dementia transition at age of diagnosis. If the subject dies before the dementia diagnosis, we consider that the health-death transition occurs at age of death and time-to-dementia is censored at age of death.

 \subsection{Results}
 \label{results}
 Table \ref{tableau::simu} displays the results of the simulation study for (a) visits every 2 years, (b) visits every 4 years, for the joint latent class model for interval-censored semi-competing events on the left and for the joint latent class model for standard competing risks on the right.\\

\begin{table}
\begin{center}
(a) Visits every 2 years
\end{center}
\setkeys{Gin}{keepaspectratio}
\resizebox*{\textwidth}{\textheight}{
\begin{tabular}{c|c }
\begin{tabular}{c c c c c c c }
\hline
&&\multicolumn{5}{c}{Joint illness-death model$^*$}\\\hline
&&$\beta$&$\hat{\beta}$& $ASE$  & $ESE$   &  Cover Rate\\\hline
Class Membership& $\zeta_1$ &0.00   &0.04   &0.2949  &  0.3208   &0.95 \\\hline

Baseline transition intensities of events&$\lambda^{(1)}_{011}$&3.20   &3.23   &0.4392 &   0.4908   &0.93 \\
&$\lambda^{(1)}_{012}$&3.50  & 3.58 & 0.3863  &  0.4333  &  0.95\\
&$\lambda^{(2)}_{011}$&0.11  & 0.11 & 0.0017  &  0.0027  &  0.96\\
&$\lambda^{(2)}_{012}$&0.10  & 0.10 & 0.0009  &  0.0009  &  0.92\\
&$\lambda^{(1)}_{021}$&3.50  & 3.54 & 0.3476  &  0.3680  &  0.94\\
&$\lambda^{(1)}_{022}$&3.40  & 3.44 & 0.2324  &  0.2493  &  0.93\\
&$\lambda^{(2)}_{021}$&0.11  & 0.11 & 0.0008  &  0.0010  &  0.93\\
&$\lambda^{(2)}_{022}$&0.10  & 0.10 & 0.0006  &  0.0006  &  0.94\\
&$\lambda^{(1)}_{121}$&2.78  & 2.85 & 0.5938  &  0.6210  &  0.92\\
&$\lambda^{(1)}_{122}$&3.14  & 3.30 & 0.6908  &  0.7144  &  0.92\\
&$\lambda^{(2)}_{121}$&0.12  & 0.12 & 0.0138  &  0.0224  &  0.84\\
&$\lambda^{(2)}_{122}$&0.11  & 0.11 & 0.0064  &  0.0091  &  0.85\\\hline
Event covariates&$\gamma_{01}$&0.02   & 0.03  &   0.2308 &   0.2323 & 0.95 \\
&$\gamma_{02}$ &0.67   &0.69 &     0.1514  &  0.1525&     0.97 \\
&$\gamma_{12}$ &0.47   &0.49&    0.2737   & 0.3104   &  0.92 \\ \hline
Latent process&$\beta_{01}$ & 30.22   &30.27&  0.7851 &   0.8325  &   0.93\\
&$\beta_{11}$ &32.96&32.98  &  0.5162 &   0.5276  &   0.95\\
&$\beta_{02}$ &-5.76    &-5.76  &  0.5678 &   0.5921  &   0.93\\
&$\beta_{12}$ &-3.53&-3.51  &  0.2029 &   0.2038  &   0.94\\
&$\beta_{X}$&0.08        & 0.03  &  0.4495 &   0.4660  &   0.94\\ \hline
Cholesky transformation&U(1,1) &4.93   &4.88 &   0.2924 &   0.2820  &  0.95\\
of the B matrix&U(1,2) &-1.15   & -1.11 &   0.2069 &   0.1992  &  0.96\\
$UU^{\top}=B$&U(2,2)&1.46   & 1.42 &   0.1392 &   0.1385  &  0.95\\ \hline
Measurement error &$\sigma_e$&3.47   &3.47&   0.0515  &  0.0534 &0.91\\ \hline
\multicolumn{7}{c}{$^*$based on 492 samples with convergence criteria fulfilled}\\

\end{tabular} &

\begin{tabular}{ c c c c c c }
\hline
\multicolumn{6}{c}{Joint competing risks model$^*$}\\\hline
&$\beta$&$\hat{\beta}$& $ASE$  & $ESE$   &  Cover Rate\\\hline
 $\zeta_1$ &0.00  &-0.19  & 0.4384 &   0.5344 & 0.91\\\hline
$\lambda^{(1)}_{011}$&3.20  &3.04 &   0.5571  &  0.6202 &  0.94\\
$\lambda^{(1)}_{012}$&3.50  &3.09  &   0.3999  &  0.4444 &  0.80\\
$\lambda^{(2)}_{011}$&0.11  &0.11  &   0.0099  &  0.0397 &  0.94\\
$\lambda^{(2)}_{012}$&0.10  &0.10  &   0.0354  &  0.0572 &  0.97\\
$\lambda^{(1)}_{021}$&3.50  &3.69  &   0.3421  &  0.3477 &  0.92\\
$\lambda^{(1)}_{022}$&3.40  &3.42  &   0.2162  &  0.2354 &  0.93\\
$\lambda^{(2)}_{021}$&0.11  &0.11  &   0.0008  &  0.0008 &  0.85\\
$\lambda^{(2)}_{022}$&0.10  &0.10  &   0.0006  &  0.0006 &  0.92\\
$\vphantom{\lambda^{(2)}_{1}}$\\
$\vphantom{\lambda^{(2)}_{1}}$\\
$\vphantom{\lambda^{(2)}_{1}}$\\
$\vphantom{\lambda^{(2)}_{1}}$\\  \hline
$\gamma_{01}$&0.02  &-0.08  & 0.2145&    0.2157& 0.93\\
$\gamma_{02}$ &0.67 & 0.66  & 0.1456&    0.1471& 0.95\\
$\vphantom{\gamma_{02}}$\\\hline
$\beta_{01}$&30.22       & 30.05 &  0.9237 &   0.9604 &0.94\\
$\beta_{11}$ &32.96  & 32.82 &  0.5106 &   0.5133 &0.94\\
$\beta_{02}$ &-5.76      & -6.04 &  0.6569 &   0.6606 &0.94\\
$\beta_{12}$ &-3.53  & -3.53 &  0.1988 &   0.1988 &0.95\\
$\beta_{X}$&0.08          &  0.06 &  0.4479 &   0.4698 &0.94\\\hline
U(1,1) &4.93  & 4.85 & 0.2994  &  0.2914& 0.94\\
U(1,2) &-1.15 &-1.10 & 0.2081  &  0.1991& 0.95\\
U(2,2)&1.46   & 1.43 & 0.1391  &  0.1389& 0.94\\\hline
$\sigma_e$&3.47  &3.47&0.0515  &  0.0535 & 0.94\\\hline
\multicolumn{6}{c}{$^*$based on 497 samples with convergence criteria fulfilled}\\
\end{tabular}
\end{tabular}}

\vspace{1cm}
\begin{center}
(b) Visits every 4 years
\end{center}

\resizebox*{\textwidth}{\textheight}{
\begin{tabular}{c c }
\begin{tabular}{ c c c c c c c }
\hline
&&\multicolumn{5}{c}{Joint illness-death model$^*$}\\\hline
&&$\beta$&$\hat{\beta}$& $ASE$  & $ESE$   &  Cover Rate\\\hline
Class Membership&  $\zeta_1$ &0.00  & 0.05&  0.3563  &  0.3961&0.95\\\hline

Baseline transition intensities of events
&$\lambda^{(1)}_{011}$&3.20  &3.24& 0.5170 &   0.5556& 0.93\\
&$\lambda^{(1)}_{012}$&3.50  &3.57& 0.4232 &   0.4842& 0.93\\
&$\lambda^{(2)}_{011}$&0.11  &0.11& 0.0033 &   0.0053& 0.96\\
&$\lambda^{(2)}_{012}$&0.10  &0.10& 0.0030 &   0.0053& 0.94\\
&$\lambda^{(1)}_{021}$&3.50  &3.52& 0.3799 &   0.3944& 0.94\\
&$\lambda^{(1)}_{022}$&3.40  &3.44& 0.2438 &   0.2649& 0.93\\
&$\lambda^{(2)}_{021}$&0.11  &0.11& 0.0010 &   0.0013& 0.94\\
&$\lambda^{(2)}_{022}$&0.10  &0.10& 0.0007 &   0.0007& 0.94\\
&$\lambda^{(1)}_{121}$&2.78  &2.90& 0.6621 &   0.7284& 0.89\\
&$\lambda^{(1)}_{122}$&3.14  &3.31& 0.7530 &   0.7818& 0.92\\
&$\lambda^{(2)}_{121}$&0.12  &0.12& 0.0233 &   0.0357& 0.81\\
&$\lambda^{(2)}_{122}$&0.11  &0.11& 0.0102 &   0.0256& 0.83\\\hline
Event covariates
&$\gamma_{01}$&0.02  &  0.04 & 0.2556   & 0.2680 &  0.94\\
&$\gamma_{02}$ &0.67  & 0.69 & 0.1595   & 0.1639 &  0.95\\
&$\gamma_{12}$ &0.47  & 0.51 & 0.3073   & 0.3658 &  0.91\\\hline
Latent process
&$\beta_{01}$& 30.22     & 30.28& 0.7995  &  0.8250& 0.94\\
&$\beta_{11}$ &32.96 & 32.99& 0.5210  &  0.5334& 0.95\\
&$\beta_{02}$ &-5.76     & -5.77& 0.5758  &  0.6007& 0.94\\
&$\beta_{12}$ &-3.53 & -3.51& 0.2046  &  0.2077& 0.94\\
&$\beta_{X}$&0.08         &  0.02 &0.4528  &  0.4766& 0.95\\\hline
Cholesky transformation&U(1,1)& 4.93&4.88& 0.2937 &   0.2844& 0.96\\
of the B matrix&U(1,2) &-1.15  &1.11& 0.2073 &   0.1994& 0.96\\
$UU^{\top}=B$&U(2,2)&1.46                   &1.42& 0.1408 &   0.1418& 0.94\\\hline
Measurement error &$\sigma_e$&3.47  &3.47&0.0515&    0.0534& 0.94\\\hline
\multicolumn{7}{c}{$^*$based on 490 samples with convergence criteria fulfilled}\\
\end{tabular}&

\begin{tabular}{ c c c c c c }
\hline
\multicolumn{6}{c}{Joint competing risks model$^*$}\\\hline
&$\beta$&$\hat{\beta}$& $ASE$  & $ESE$   &  Cover Rate\\\hline
 $\zeta_1$ &0.00  &-0.33&   0.5909 &   0.6620  & 0.82\\\hline
$\lambda^{(1)}_{011}$&3.20  & 2.85&   0.8301  &  1.1372 &0.88\\
$\lambda^{(1)}_{012}$&3.50  & 2.69&   0.4083  &  0.4822 &0.48\\
$\lambda^{(2)}_{011}$&0.11  & 0.12&   0.0410  &  0.0839 &0.83\\
$\lambda^{(2)}_{012}$&0.10  & 0.10&   0.0111  &  0.0391 &0.96\\
$\lambda^{(1)}_{021}$&3.50  & 3.74&   0.3412  &  0.3618 &0.89\\
$\lambda^{(1)}_{022}$&3.40  & 3.41&   0.2106  &  0.2300 &0.91\\
$\lambda^{(2)}_{021}$&0.11  & 0.11&   0.0008  &  0.0008 &0.63\\
$\lambda^{(2)}_{022}$&0.10  & 0.10&   0.0006  &  0.0007 &0.82\\
$\vphantom{\lambda^{(2)}_{1}}$\\
$\vphantom{\lambda^{(2)}_{1}}$\\
$\vphantom{\lambda^{(2)}_{1}}$\\
$\vphantom{\lambda^{(2)}_{1}}$\\  \hline
$\gamma_{01}$&0.02  &-0.18&  0.2190  &  0.2282 & 0.85\\
$\gamma_{02}$ &0.67  &0.65&   0.1445   & 0.1486  & 0.94\\
$\vphantom{\gamma_{02}}$\\ \hline
$\beta_{01}$&30.22      &29.44 & 1.0022  &  1.0924 & 0.94\\
$\beta_{11}$ &32.96 &32.68 & 0.5026  &  0.5193 & 0.92\\
$\beta_{02}$ &-5.76     &-6.20 & 0.7162  &  0.7353 & 0.93\\
$\beta_{12}$ &-3.53 &-3.53 & 0.1957  &  0.1976 & 0.94\\
$\beta_{X}$&0.08         & 0.10 & 0.4454  &  0.4799 & 0.92\\\hline
U(1,1) &4.93   &  4.85&0.3010  &  0.2963 & 0.94\\
U(1,2) &-1.15  & -1.09& 0.2082 &   0.1991&  0.95\\
U(2,2)& 1.46   &  1.44& 0.1388 &   0.1393&  0.95\\\hline
$\sigma_e$&3.47   &3.47&   0.0515 &   0.0534&0.94 \\ \hline
\multicolumn{6}{c}{$^*$based on 490 samples with convergence criteria fulfilled}\\
\end{tabular}
\end{tabular}}

\vspace{1cm}

\caption{Results of the simulation study comparing estimates of the joint latent class markovian illness-death model for interval-censored events and the joint latent class competing risks model. A total of 500 samples of 500 subjects were generated with a joint markovian illness-death model with visits every 2 or 4 years. ASE is the asymptotic standard error, ESE is the empirical standard error and the coverage rate is calculated from the 95\% confidence interval.}
\label{tableau::simu}

\end{table}

The top left part of Table  \ref{tableau::simu} shows small biases and good coverage rates of the 95\% confidence interval for the 25 parameters except for the two scale parameters of the dementia-death transition, $\lambda^{(2)}_{121}$ and $\lambda^{(2)}_{122}$, which have lower coverage rates because their standard errors are under-estimated. This may be due to the small number of observed transitions from dementia to death, in these 500-subject samples. Indeed, the simulations made on 1000 subjects and presented in Web Table 1 show better coverage rates. \\

The simulations were also carried out with visit intervals of four years. We can see on the bottom left part of Table \ref{tableau::simu} that the parameters are still well estimated but have higher variances, especially for the illness-death parameter estimators, as the number of unobserved transitions increases when the censoring interval gets bigger.\\

When compared with the competing risks estimations on the top right part of Table \ref{tableau::simu}, we observe higher biases for the shape parameters for the transition toward dementia in both classes, $\lambda^{(1)}_{011}$ and $\lambda^{(1)}_{012}$, and  toward death $\lambda^{(1)}_{021}$ and an under-estimation of the standard errors of the four parameters for the transition toward dementia. These trends are more pronounced for the 4-year-visit-interval data (see part b in Table  \ref{tableau::simu}), leading to poor coverage rates that worsen further when the sample size increases due to smaller standard errors (see Web Table 1 for N=1000).\\


We also assessed the semi-markovian model, with visit intervals of two and four years. The estimates have small biases and good coverage rates in the longitudinal and the illness-death parts (see Web Table 2).

\section{Application}
The joint latent class illness-death model was applied to a French prospective cohort, the Paquid cohort, to distinguish different profiles of cognitive decline in the elderly associated with the transition intensity of Alzheimer's disease and death. We compared markovian and semi-markovian models, in order to  determine whether the transition intensity of death among demented subjects depended more on age than on duration of dementia.
\subsection{Data}
The Paquid cohort \citep{letenneur} involves 3,777 subjects from two French administrative departments, Dordogne and Gironde. The subjects were 65 years old or older at entry and they were visited every 2 or 3 years at home to undergo a battery of psychometric tests. The diagnosis of dementia was based on a two-phase screening procedure according to DSM III R criteria for dementia \citep{DSM}. In this work, we focused on the ISAACS set test, scored from 0 to 40, assessing verbal fluency. Subjects had to produce up to 10 words from four different semantic categories within 15 seconds for each category. \\

We selected subjects who were healthy at inclusion and who completed at least one ISAACS set test (until their diagnosis for subjects diagnosed as demented). Among the 3,777 subjects of the initial sample, 102 were excluded because they were prevalent cases at the first visit and 150 were excluded because they had completed no tests during the follow-up. Finally, the sample under study involved 3,525 subjects, including 57.8\% of women and 65.9\% of subjects with a low level of education. A total of 23.8\% of subjects were diagnosed as demented, including 19.8\% who died during the follow-up, and 65.1\% who died before the dementia diagnosis. 

\subsection{Comparison of models}
The joint markovian model and the joint semi-markovian model were estimated and compared on the Paquid cohort. In both models, the sub-model for the longitudinal marker assumed a quadratic trend with three class-specific fixed and random effects of time and common effects for gender ($Sex$ = 0 for men, 1 for women) and educational level ($Educ$ = 1 for subjects who obtained their primary school diploma, 0 for others). Given class $g$, the mixed model was defined by:
\begin{equation*}
\begin{split}
\Lambda_i(t)=&\beta_{0g}+\beta_{1g}~t+\beta_{2g}~\frac{t^2}{10}+\beta_{3}~Educ_i+\beta_{4}~Educ_i \times t\\
&+\beta_5~Educ_i \times \frac{t^2}{10}+\beta_{6}~Sex_i+u^{(0)}_{ig}+u^{(1)}_{ig}t+u^{(2)}_{ig}\frac{t^2}{10}, 
\end{split}
\end{equation*}
where $u_{ig}=(u^{(0)}_{ig},u^{(1)}_{ig},u^{(2)}_{ig})^{\top} \sim \mathcal{N}(0,\sigma^2_gB)$. Time is a linear transformation of age : $t=\frac{age-65}{10}$. The interaction between time and gender was not accounted for because it appeared insignificant in previous analyses. A Beta cumulative distribution function was used to link the observed scores to the latent process:
\begin{equation}
\tilde{Y}_{ij}=\Psi(Y_{ij};\eta^{(1)},\eta^{(2)},\eta^{(3)},\eta^{(4)})=\frac{B \big(Y_{ij};\eta^{(1)},\eta^{(2)} \big)-\eta^{(3)}}{\eta^{(4)}} = \Lambda_{i}(t_{ij})+\epsilon_{ij},
\end{equation}
where $\epsilon_{ij} \sim  \mathcal{N}(0,\sigma^2_e)$. The models for the transition intensities were proportional hazards models with common effects of the two covariates and class-specific Weibull baseline transition intensities, as functions of age. We compared the markovian model defined by: 
\begin{equation*}
\alpha_{klig}(t)=\alpha_{klg}^0(t)~\rm{\exp}\big(\gamma^s_{kl}~Sex_i+\gamma^e_{kl}~Educ_i\big),~ k \in [0,1], ~ l \in [1,2],
\end{equation*}

and the semi-markovian model where the third transition intensity had the following form:
\begin{equation*}
\alpha_{12ig}(t,T^A_i)=\alpha_{12g}^0(t-T^A_i)~\rm{\exp}\big(\gamma^s_{12}~Sex_i+\gamma^e_{12}~Educ_i\big),
\end{equation*}
where $T^A$ is the age at dementia onset.

\begin{table}[h!]
\begin{center}
\begin{tabular}{ c c c}

&\multicolumn{2}{c}{BIC}\\
\hline
&Markovian&Semi-markovian\\\hline
G=1& 106928 &  107050\\
G=2&106315 &  106368\\
G=3&106120  &106185\\
G=4& 106058 &106113 \\
 G=5&106091  &106099 \\

\hline
\end{tabular}
\end{center}
\vspace{0.5cm}
\caption{Comparison of BIC of markovian and semi-markovian joint latent class illness-death models, with a total number of classes varying from 1 to 5 (Paquid, N=3,525).}
\label{table::comp}
\end{table}

Table \ref{table::comp} shows that the markovian model fits the data better, with smaller BIC values irrespective of the number of classes. Thus, the transition intensity of death among demented subjects depends more on age than on dementia duration.\\


The minimum value of BIC was obtained with G = 4 classes within the markovian framework. Thus, this model will be detailed below. Note that a higher degree of heterogeneity is expected within the semi-markovian framework, since the minimum value of BIC was obtained with five classes, but the fit was not as good as the four-class markovian model. 

\subsection{Results}
\label{appli::results}
\begin{figure}
\begin{center}
\includegraphics[width=5cm]{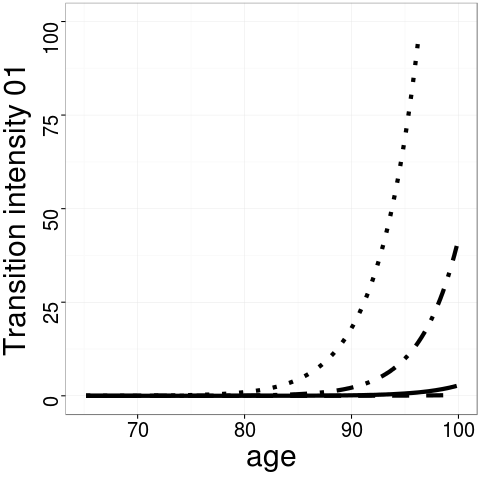}
\includegraphics[width=5cm]{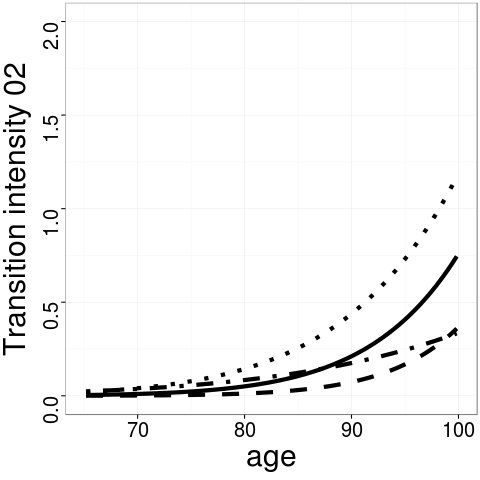}
\includegraphics[width=5cm]{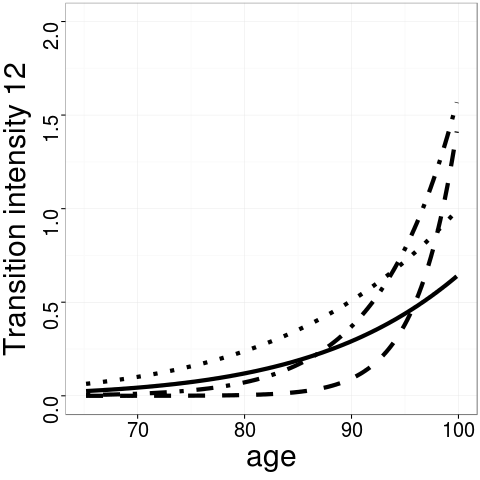}\\
A) Class-specific transition intensities of the illness-death model for men with a low level of education.\\
\vspace{0.5cm}
\includegraphics[width=5cm]{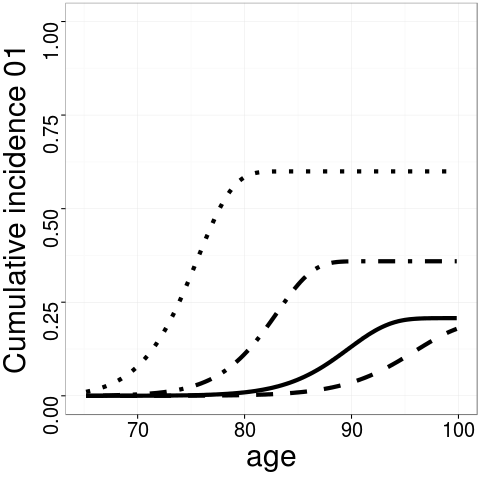}
\includegraphics[width=5cm]{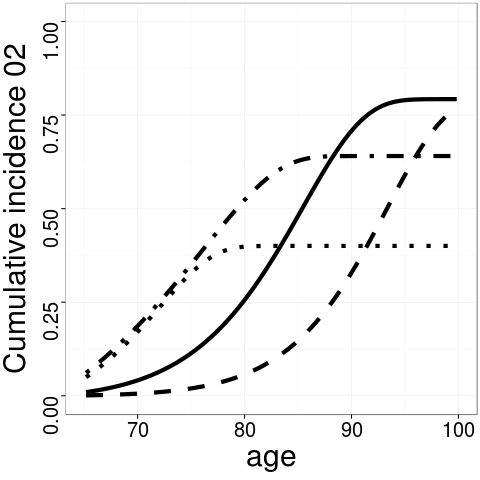}
\includegraphics[width=5cm]{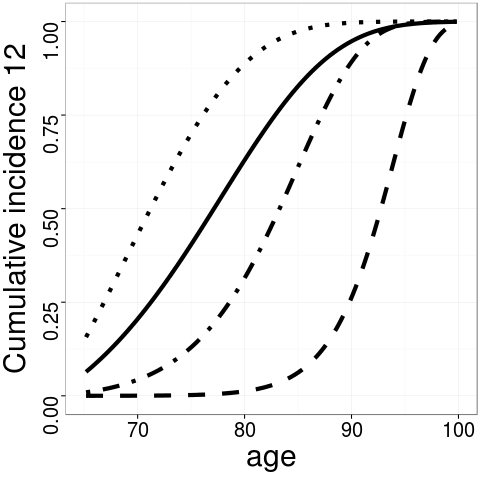}\\
B) Class-specific cumulative incidences of the illness-death model for men with a low level of education.\\
\vspace{0.5cm}
\includegraphics[width=7cm]{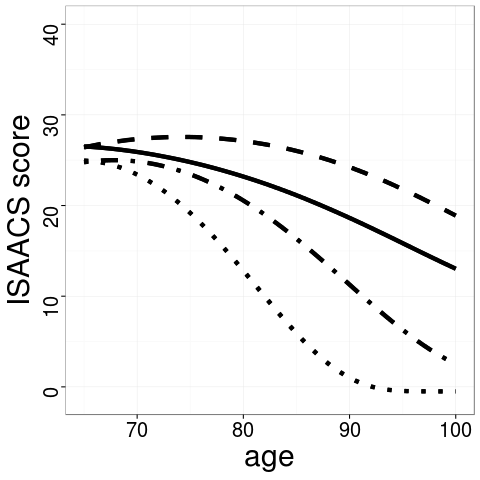}\\
C) Class-specific ISAACS set test score trajectories for men with a low level of education.
\end{center}
\caption{Class-specific estimated transition intensities, cumulative incidences and mean longitudinal trajectories of the latent class illness-death model for each class (class 1: dashed line, class 2: dotted line, class 3: dotdashed line, class 4: solid line).}
\label{graph::plots}
\end{figure}

Figure \ref{graph::plots} represents the three transition intensities (Part A), the three cumulative incidences (Part B) and the estimated mean ISAACS trajectories (Part C) for men with a low level of education (with no primary school diploma), in each class. The first class includes 7.3\% of the sample and will be denoted as the `healthy class' as it has the lowest transition intensities of dementia and  death until advanced ages, as well as the slightest mean decline of the ISAACS set test score (Figure \ref{graph::plots} C). The cumulative incidences show that most dementia and death occurrences arise after age 85 in this class (Figure \ref{graph::plots} B). At the other extreme, the second class, including 8.0\% of the population, will be denoted the `high transition intensities class' since it has the highest transition intensities of dementia and death for healthy people and the fastest and deepest decline of ISAACS scores.  As shown in Figure \ref{graph::plots} B, most of these subjects are demented or dead before age 80. 
The third class, accounting for 34.2\% of the population, also has a high transition intensity of dementia and a steep cognitive decline, but these occur at later ages than in the second class, and a medium transition intensity of death for healthy people. The cumulative incidences show that about 60\% of this group die without dementia before age 85 and 30\% become demented before age 90. Finally, the fourth class includes 50.5\% of the population and is quite similar to the first one, with more pronounced cognitive decline and higher transition intensities of death among both healthy and demented subjects. About 80\% die without dementia and 20\% become demented before age 95 in this class whereas these figures are only  reached 5 to 10 years later in the first class.\\


Estimates of the main parameters of the retained model are presented in Web Table 3. Educational level is associated with lower transition intensities of dementia and death among healthy people but not with the transition intensity of death among demented people, after adjusting for the heterogeneity of the transition intensities due to the classes. Gender is not associated with the transition intensity of dementia but healthy and demented females have a lower transition intensity of death. Note that these estimates do not represent the global effect of gender and education on the population since the proportion of women and educated subjects is quite different between classes (see Web Table 4).

\subsection{Goodness-of-fit}
\label{goodness}
As loss of follow-up may be linked to a change in the cognitive test \citep{MCAR}, missing data are not missing completely at random; so we assessed the goodness-of-fit conditionally on random effects. Web Figure 1 displays the predicted weighted mean of ISAACS scores given the random effects and the classes, and the weighted mean of the observed scores for each class. The predicted mean is close to the observed mean and is within the confidence interval which increases over time as there is less data. The estimated class-specific cumulative incidences averaged on the covariates on part B of Web Figure 1 are compared to the estimations, also averaged on the covariates, obtained by a semi-parametric illness-death model with baseline transition intensities modelled by M-splines, estimated with a weighted penalized likelihood. Here again, the estimations are close and the graphs show that the model fits the data well.

\subsection{Posterior classification}
\label{posterior}
Considering each posterior class, we compute the mean probability of belonging to each of the four classes in order to quantify the discriminatory ability of the model (see Web Table 5). For each class, the probability of belonging to the allocated class is above 63\% and the mean probability of belonging to another class is less than 10\%, which is quite satisfactory.

\subsection{Post-fit trajectories}
\label{postfit}
\begin{figure}
  \begin{center}
    \includegraphics[width=10cm]{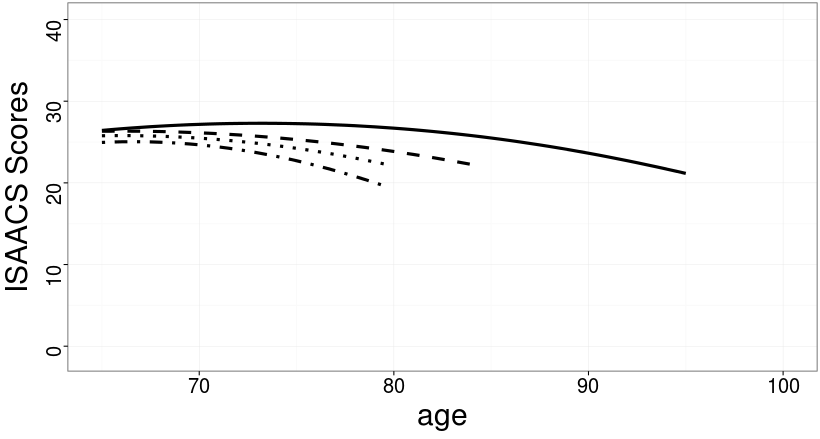}
    \end{center}
    \caption{Predicted ISAACS trajectories for a man with a low level of education, alive and healthy at age 95 (solid line) or 85 (dashed line), a man who dies at age 80 without dementia (dotted line) and a man who becomes demented at age 80 (dotdashed line). The trajectory is plotted from 65 years old until the age at dementia diagnosis, loss of follow-up or death.}

      \label{postfit_traj}
\end{figure}

It is of interest to estimate the typical cognitive decline of subjects who were demented or deceased at a given age, as well as the evolution of subjects who were alive and non-demented at an advanced age. Thus, we computed the mean trajectories of the ISAACS scores for a man with a low level of education for 4 different cases: alive and healthy at age 95 or at 85, dead without dementia at age 80 and a man with dementia onset at age 80. The expectation for the first case is given by

\begin{equation}
 E(Y(t)|T^A_i> 95, T^D_i>95;\hat{\theta}_G)= \sum_{g=1}^G E(Y(t)|c_i=g;\hat{\theta}_G) P(c_i=g|T^A_i> 95, T^D_i>95;\hat{\theta}_G),
\end{equation}
and the other cases are computed the same way. As expected, we observe on Figure \ref{postfit_traj} that the trajectories of the men alive and healthy are the highest ones with the smallest declines (solid and dashed lines). Nevertheless, these estimates  show a slight decline of the ISAACS scores in older ages among healthy people, probably due to a slowing of cognitive processes. The decline of the man who dies at age 80 without dementia is slightly more pronounced, highlighting the so-called terminal decline before death, while the decline of the man who becomes demented at age 80 is deeper. The `terminal decline'  has been described by other authors \citep{terminal_decline}, however, it has not been distinguished from the decline toward dementia. As expected, the decline is deeper in the pre-dementia phase (dotdashed curves). 

\section{Discussion}
We proposed a joint latent class illness-death model for semi-competing interval-censored events and longitudinal data. Joint models have previously been developed to capture the correlation between a longitudinal marker and competing risks but no previous model has accounted for interval censoring \citep{Elashoff, Williamson,Proust_2015}. Our simulations highlighted biased estimates of the joint model for standard competing risks. By dealing with interval censoring, the proposed method corrects for this bias and highlights different profiles of cognitive decline associated with different transition intensities of death and dementia. Subsequently,  the mean trajectories of cognitive decline for people who were demented or deceased at a given age can be estimated. This makes it possible to distinguish the cognitive decline of the healthy elderly from the cognitive decline before death without dementia and the cognitive decline in the pre-dementia phase that appears to be the steepest.\\

We chose a latent class approach to account for the heterogeneity of cognitive aging. It could be also interesting to develop a joint model for semi-competing interval-censored events and longitudinal data in a shared random effects framework. Nevertheless, with some realistic assumptions, handling of interval censoring is not as critical in shared random effects models because the transition intensity of death may depend on the current value of the marker and possibly on the current slope. Thus, conditionally on the current value and the slope of the marker (which is the cognitive test in our application), it may be sensible to assume that the transition intensity of death is identical for demented and healthy subjects. In this case, as shown in Dantan et al. (2011), it is not necessary to distinguish possible individual trajectories between the last visit without dementia and the end of follow-up or death. The likelihood for interval-censored data is identical to the one without interval censoring, as long as we impute the middle of the censoring interval for subjects diagnosed as demented. Moreover, \cite{karen} showed that the uncertainty regarding time to dementia onset, among the subjects diagnosed as demented, is not a major issue as long as the intervals between visits are short enough. Nevertheless, extending the joint shared random effects model to include an illness-death submodel instead of a standard times-to-events model could be useful to study the association between  marker evolution and transition intensity of death among diseased subjects.\\

A score test was previously developed \citep{score_test} to assess the assumption of independence between the marker and the time to an event conditionally on the latent classes. This score was extended to the framework of competing risks and multiple markers by \cite{Proust_2015}. In the future, an extension to joint illness-death model accounting for interval censoring could be useful.
\backmatter
\section*{Acknowledgements}
We thank Florian Arnoux for his participation in programming. Computer time for this study was provided by the computing facilities MCIA (Mésocentre de Calcul Intensif Aquitain) of the Université de Bordeaux and of the Université de Pau et des Pays de l'Adour. Anaïs Rouanet was funded by an INSERM/Region Aquitaine PhD allocation and the Paquid study is funded by IPSEN and Novartis laboratories and the Caisse Nationale de Solidarité et d'Autonomie. 
\section*{Supplementary Material}

Web Appendices, Tables, and Figures referenced in Sections \ref{results}, \ref{appli::results}, \ref{goodness} and \ref{posterior} are available with this paper at the Biometrics website on Wiley Online Library.



\label{lastpage}


\begin{thebibliography}{}

\bibitem[\protect\citeauthoryear{American Psychiatric Association}{1987}]{DSM}
American Psychiatric Association (1987). \textit{Diagnostic and Statistical Manual of Mental Disorders.} (DSM-III-R. 3rd edition revised).


\bibitem[\protect\citeauthoryear{Dantan, Joly and Jacqmin-Gadda}{2011}]{Dantan}
Dantan, E., Joly, P., Dartigues, J.-F., and Jacqmin-Gadda, H. (2011). Joint model with latent state for longitudinal and multistate data. {\it Biostatistics} {\bf 12}, 723--736.

\bibitem[\protect\citeauthoryear{Elashoff et al.}{2008}]{Elashoff}
Elashoff, R. M., Li, G., and Li, N. (2008). A joint model for longitudinal measurements and survival data in the presence of multiple failure types. {\it Biometrics} {\bf 64}, 762--71.
\bibitem[\protect\citeauthoryear{Hawkins, Allen and Stromberg}{2001}]{Hawkins}
Hawkins, D. S., Allen, D. M., and Stromberg, A. J. (2001). Determining the number of components in mixtures of linear models. {\it Computational Statistics \& Data Analysis} {\bf 38}, 15--48.


\bibitem[\protect\citeauthoryear{Jacqmin-Gadda et al.}{1997}]{MCAR}
Jacqmin-Gadda, H., Fabrigoule, C., Commenges, D., and Dartigues, J.-F. (1997). A five year longitudinal study of mini mental state examination in normal aging. {\it American Journal of Epidemiology} {\bf 145}, 498--506.

\bibitem[\protect\citeauthoryear{Jacqmin-Gadda et al.}{2010}]{score_test}
Jacqmin-Gadda, H., Proust-Lima, C., Taylor, J., and Commenges, D. (2010). Score Test for Conditional Independence Between Longitudinal Outcome and Time to Event Given the Classes in the Joint Latent Class Model. {\it Biometrics} {\bf 66}, 11--19.

\bibitem[\protect\citeauthoryear{Joly et al.}{2002}]{joly_2002}
Joly, P., Commenges, D., Helmer, C., and Letenneur, L. (2002). A penalized likelihood approach for an illness–death model with interval-censored data: application to age-specific incidence of dementia. {\it Biostatistics} {\bf 3}, 433--443.


\bibitem[\protect\citeauthoryear{Leffondre et al.}{2013}]{karen}
Leffondre, K., Touraine, C., Helmer, C., and Joly, P. (2003). Interval-censored time-to-event and competing risk with death: is the illness-death model more
accurate than the Cox model? {\it International Journal of Epidemiology} {\bf 42}, 1177--1186.

\bibitem[\protect\citeauthoryear{Letenneur et al.}{1994}]{letenneur}
Letenneur, L., Commenges, D., Dartigues, J.-F., and Barberger-Gateau, P. (1994). Incidence of dementia and Alzheimer’s disease in elderly community residents of south-western France. {\it International Journal of Epidemiology} {\bf 23}, 1256--1261.

\bibitem[\protect\citeauthoryear{Lin, Turnbull and Slate}{2002}]{Lin}
Lin, H., Turnbull, B. W., McCulloch, C. E., and Slate, E.H. (2002). Latent class models for joint analysis of longitudinal biomarker and event process data: application to longitudinal prostate-specific antigen readings and prostate cancer. {\it Journal of the American Statistical association} {\bf 97}, 53--65.

\bibitem[\protect\citeauthoryear{Marquardt}{1963}]{marq98}
Marquardt, D. (1963). An algorithm for least-squares estimation of nonlinear parameters. {\it Journal of Applied Mathematics} {\bf 11}, 431--441.

\bibitem[\protect\citeauthoryear{Proust-Lima et al.}{2009}]{CSDA}
Proust-Lima, C., Joly, P., Dartigues, J.-F., and Jacqmin-Gadda, H. (2009). Joint modelling of multivariate longitudinal outcomes and a time-to-event: A nonlinear latent class approach. {\it Computational Statistics and Data Analysis} {\bf 53}, 1142--1154.

\bibitem[\protect\citeauthoryear{Proust-Lima and Jacqmin-Gadda}{2015, submitted}]{Proust_2015}
Proust-Lima, C. and Jacqmin-Gadda, H. (2015). Joint modelling of repeated multivariate cognitive measures and competing risks of dementia and death: a latent process and latent class approach {\it submitted}.

\bibitem[\protect\citeauthoryear{Rizopoulos}{2012}]{rizopoulos}
Rizopoulos, D. (2012). {\it Joint Models for Longitudinal and Time-to-event Data: With Applications in R}, Vol. 6. Boca Raton: Chapman \& Hall.

\bibitem[\protect\citeauthoryear{Schwartz}{1978}]{BIC}
Schwartz, G. (1978). Estimating the dimension of a model. {\it The Annals of Statistics} {\bf 6}, 461--464.


\bibitem[\protect\citeauthoryear{Touraine, Helmer and Joly}{2013}]{touraine}
Touraine, C., Helmer, C., and Joly, P. (2013). Predictions in an illness-death model. {\it Statistical Methods in Medical Research} DOI: 10.1177/0962280213489234 


\bibitem[\protect\citeauthoryear{Tsiatis and Davidian}{2004}]{tsiatis}
Tsiatis, A. A. and Davidian, M. (1997). Joint modeling of longitudinal and time-to-event data : An overview. {\it Statistica Sinica} {\bf 14}, 809--834.


\bibitem[\protect\citeauthoryear{Williamson et al.}{2008}]{Williamson}
Williamson, P. R., Kolamunnage-Dona, R., Philipson, P., and Marson, A. G. (2008). Joint modelling of longitudinal and competing risks data. {\it Statistics in medicine} {\bf 27}, 6426--6438.

\bibitem[\protect\citeauthoryear{Wilson et al.}{2003}]{terminal_decline}
Wilson, R. S., Beckett, L. A., Bienias, J. L., Evans, D. A., and Bennett, D. A. (2003). Terminal decline in cognitive function. {\it Neurology} {\bf 60}, 1781--1787.


\bibitem[\protect\citeauthoryear{Wulfsohn and Tsiatis}{1997}]{Wulfsohn}
Wulfsohn, M. S. and Tsiatis, A. A. (1997). A joint model for survival and longitudinal data measured with error. {\it Biometrics} {\bf 53}, 330--339.



\end{thebibliography}
\end{document}